\documentclass[12pt]{article}
\usepackage{graphicx}
\def\hybrid{\topmargin 0pt      \oddsidemargin 0pt
        \headheight 0pt \headsep 0pt
        \voffset=-0.5cm
        \textwidth 6.25in       
        \textheight 9.5in       
        \marginparwidth 0.0in
        \parskip 5pt plus 1pt   \jot = 1.5ex}
\catcode`\@=11
\def\marginnote#1{}

\newcount\hour
\newcount\minute
\newtoks\amorpm
\hour=\time\divide\hour by60
\minute=\time{\multiply\hour by60 \global\advance\minute by-\hour}
\edef\standardtime{{\ifnum\hour<12 \global\amorpm={am}%
        \else\global\amorpm={pm}\advance\hour by-12 \fi
        \ifnum\hour=0 \hour=12 \fi
        \number\hour:\ifnum\minute<10 0\fi\number\minute\the\amorpm}}
\edef\militarytime{\number\hour:\ifnum\minute<10 0\fi\number\minute}

\def\draftlabel#1{{\@bsphack\if@filesw {\let\thepage\relax
   \xdef\@gtempa{\write\@auxout{\string
      \newlabel{#1}{{\@currentlabel}{\thepage}}}}}\@gtempa
   \if@nobreak \ifvmode\nobreak\fi\fi\fi\@esphack}
        \gdef\@eqnlabel{#1}}
\def\@eqnlabel{}
\def\@vacuum{}
\def\draftmarginnote#1{\marginpar{\raggedright\scriptsize\tt#1}}
\def\draftlabel#1{{\@bsphack\if@filesw {\let\thepage\relax
   \xdef\@gtempa{\write\@auxout{\string
      \newlabel{#1}{{\@currentlabel}{\thepage}}}}}\@gtempa
   \if@nobreak \ifvmode\nobreak\fi\fi\fi\@esphack}
        \gdef\@eqnlabel{#1}}
\def\@eqnlabel{}
\def\@vacuum{}
\def\draftmarginnote#1{\marginpar{\raggedright\scriptsize\tt#1}}

\def\draft{\oddsidemargin -.5truein
        \def\@oddfoot{\sl preliminary draft \hfil
        \rm\thepage\hfil\sl\today\quad\militarytime}
        \let\@evenfoot\@oddfoot \overfullrule 3pt
        \let\label=\draftlabel
        \let\marginnote=\draftmarginnote
   \def\@eqnnum{(\theequation)\rlap{\kern\marginparsep\tt\@eqnlabel}%
\global\let\@eqnlabel\@vacuum}  }

\def\numberbysection{\@addtoreset{equation}{section}
        \def\theequation{\thesection.\arabic{equation}}}

\def\underline#1{\relax\ifmmode\@@underline#1\else
        $\@@underline{\hbox{#1}}$\relax\fi}

\def\titlepage{\@restonecolfalse\if@twocolumn\@restonecoltrue\onecolumn
     \else \newpage \fi \thispagestyle{empty}\c@page\z@
        \def\thefootnote{\fnsymbol{footnote}} }

\def\endtitlepage{\if@restonecol\twocolumn \else  \fi
        \def\thefootnote{\arabic{footnote}}
        \setcounter{footnote}{0}}  
\relax

\numberbysection
\hybrid

\newfont{\Bbb}{msbm10 scaled 1\@ptsize00}

\newcommand{\NN}{\mbox{\Bbb N}}

\newcommand{\ZZ}{\mbox{\Bbb Z}}
\newfont{\Bbbb}{msbm7 scaled 1\@ptsize00}
\newcommand{\cc}{\raise-1pt\hbox{$\mbox{\Bbbb C}$}}
\newcommand{\zz}{\raise-1pt\hbox{$\mbox{\Bbbb Z}$}}

\newtheorem{theorem}{Theorem}[section]
\newtheorem{lemma}{Lemma}[section]
\newtheorem{lemma-definition}{Lemma-Definition}[section]

\newtheorem{proposition}{Proposition}[section]

\def\beq{\begin{equation}}
\def\eeq{\end{equation}}
\def\p{\partial}

\def\res{\mathop{\hbox{res}}\limits}

\def\square{\hfill
{\vrule height6pt width6pt depth1pt} \break \vspace{.01cm}}

\begin{document}

\begin{titlepage}

\title{On integrability of the deformed Ruijsenaars-Schneider system}

\author{
A.~Zabrodin\thanks{
Skolkovo Institute of Science and Technology, Moscow 143026, 
Russia and
National Research University Higher School of Economics,
20 Myasnitskaya Ulitsa,
Moscow 101000, Russia and
NRC ``Kurchatov Institute'', Moscow, Russia;
e-mail: zabrodin@itep.ru}}

\date{December 2022}
\maketitle

\vspace{-7cm} \centerline{ \hfill ITEP-TH-28/22}\vspace{7cm}

\hfill
{\it Dedicated to the memory of Igor Krichever}

\vspace{0.5cm}

\begin{abstract}

We find integrals of motion for the recently introduced deformed
Ruijsenaars-Schneider many-body system which is the dynamical system
for poles of elliptic solutions to the Toda lattice with constraint
of type B. Our method is based on the fact that equations of motion
for this system coincide with those for pairs of Ruijsenaars-Schneider
particles which stick together preserving a special fixed distance between
the particles.

\end{abstract}


\end{titlepage}

\tableofcontents

\vspace{5mm}

\section{Introduction}

Integrable many-body systems of classical mechanics play a significant
role in modern mathematical physics. They are interesting and meaningful
from both mathematical and physical points of view and have important
applications and deep connections with different problems in mathematics
and physics. The history of integrable many-body systems starts from the
famous Calogero-Moser (CM) model \cite{Calogero71}-\cite{OP81}
which exists in rational, trigonometric or hyperbolic and elliptic versions.
In the most general elliptic case the equations of motion for the 
$N$-body CM system are
\beq\label{int1}
\ddot x_i =4\sum_{j\neq i}^N\wp '(x_{ij}), \qquad x_{ij}=x_i-x_j,
\eeq
where dot means the time derivative. Throughout the paper, we use the 
standard Weierstrass $\sigma$-, $\zeta$- and $\wp$-functions 
$\sigma (x)$, $\zeta (x)=\sigma ' (x)/\sigma (x)$
and $\wp (x)=-\zeta '(x)$ (see Appendix A for their definition and 
properties). Degenerating the elliptic functions to trigonometric and
rational ones, one obtains the trigonometric and rational versions
of the CM model. 
The elliptic CM model is Hamiltonian and completely integrable,
i.e., it has $N$ independent 
integrals of motion in involution. Integrability of the
model was proved by different methods in \cite{Perelomov}
and \cite{W77}, see also the book 
\cite{Perelomov-book}.

Later it was discovered \cite{RS86,Ruij87} that there exists a
one-parametric deformation of the CM system preserving integrability,
often referred to as relativistic extension. The parameter of the
deformation, $\eta$, in this interpretation is the inverse velocity
of light. This model is now called the Ruijsenaars-Schneider (RS)
system. Again, in its most general version the interaction 
between particles is described
by elliptic functions. The equations of motion are
\beq\label{int2}
\ddot x_i +\sum_{j\neq i}^N \dot x_i \dot x_j
\Bigl (\zeta (x_{ij}+\eta )+
\zeta (x_{ij}-\eta )-2\zeta (x_{ij})\Bigr )=0.
\eeq
A properly taken limit $\eta \to 0$ leads to equations (\ref{int1}). 
The RS system is Hamiltonian with the Hamiltonian
\beq\label{int3}
{\sf H}_1=\sum_{i=1}^N e^{p_i}\prod_{j\neq i}^N 
\frac{\sigma (x_{ij}+\eta )}{\sigma (x_{ij})}.
\eeq
Integrability of the RS system was proved in \cite{Ruij87}. 
It has conserved quantities 
${\sf H}_k$, $\bar {\sf H}_k$ $k\in \NN$, 
which are higher Hamiltonians in involution (for the $N$-particle
system the first $N$ of them are independent).

Since the seminal works \cite{AMM77}-\cite{Krichever80} it
became a common knowledge that the integrable 
many-body systems of Calogero-Moser type 
describe dynamics of poles of singular solutions (in general, elliptic
solutions) to nonlinear integrable differential equations such as 
Korteveg-de Vries (KdV) and Kadomtsev-Petviashvili (KP) equations.
In \cite{KZ95} it was shown that the RS system plays the same 
role for singular solutions to the Toda lattice equation which can be 
thought of as an integrable difference deformation of the KP equation.
(On the Toda lattice side, the parameter $\eta$ can be identified with
the lattice spacing.) Namely, the time evolution of poles in the time
$t=t_1$ of the Toda hierarchy coincides with the RS dynamics 
according to the
equations of motion (\ref{int2}). 
Later this correspondence 
was extended \cite{PZ21} to the level of hierarchies: the evolution of poles 
in the higher times $t_k$ and $\bar t_k$ 
of the Toda hierarchy was shown
to be given by the RS Hamiltonian flows with the higher Hamiltonians
${\sf H}_k$ and $\bar {\sf H}_k$. 

Recently, a deformation of the RS model was introduced \cite{KZ22}
as a dynamical system describing time evolution of poles of 
elliptic solutions
to the Toda lattice with the constraint of type B \cite{KZ22a}.
Equations of motion of the deformed RS system are
\beq\label{int4}
\ddot x_i +\sum_{j\neq i}^N \dot x_i \dot x_j
\Bigl (\zeta (x_{ij}+\eta )+
\zeta (x_{ij}-\eta )-2\zeta (x_{ij})\Bigr )+g(U_i^--U_i^+)=0,
\eeq
where
\beq\label{int5}
U_i^{\pm}=\prod_{j\neq i}^N U^{\pm}(x_{ij}), \qquad
U^{\pm}(x_{ij})=
\frac{\sigma (x_{ij}\pm 2\eta )
\sigma (x_{ij}\mp \eta )}{\sigma (x_{ij}\pm \eta )\sigma (x_{ij})}
\eeq
and $g$ is the deformation parameter. At $g=0$ we have the RS system.
It is evident that $g\neq 0$ can be eliminated from the 
formulas by re-scaling of the time variable $t\to g^{-1/2}t$. In what
follows we fix $g$ to be $g=\sigma (2\eta )$ without loss of generality.
With this choice of $g$, equations (\ref{int4}) are exactly the same 
as they appear as the dynamical equations for poles with the convention
on the choice of the time variable adopted 
in the Toda lattice with the constraint
of type B. 
In \cite{KZ22} it was shown that the $\eta \to 0$ limit of 
equations (\ref{int4})
reproduces the equations of motion 
\beq\label{int4a}
\ddot x_i+6 \sum_{j\neq i}^N(\dot x_i+\dot x_j)\wp '(x_{ij})-
72 \sum_{j,k\neq i, j\neq k}
\wp (x_{ij})\wp '(x_{ik})=0
\eeq
obtained in \cite{RZ20} for dynamics of
poles of elliptic solutions to the B-version of the KP equation (BKP). 

In \cite{KZ22} it was also shown that that the system (\ref{int4})
can be obtained by restriction of the Hamiltonian flow with the 
Hamiltonian ${\sf H}_1^- ={\sf H}_1 -\bar {\sf H}_1$ of the 
$N=2N_0$-particle RS system to the half-dimensional subspace
${\cal P}\subset {\cal F}$ of the $4N_0$-dimensional 
phase space ${\cal F}$ 
corresponding to the configurations in which the $2N_0$ particles
stick together joining in $N_0$ pairs such that the distance between
particles in each pair is equal to $\eta$. Such configurations are
immediately destroyed by the flow with the 
Hamiltonian ${\sf H}_1^+ ={\sf H}_1 +\bar {\sf H}_1$ but are
preserved by the flow with the 
Hamiltonian ${\sf H}_1^- ={\sf H}_1 -\bar {\sf H}_1$ and the 
corresponding dynamics can be restricted to the subspace ${\cal P}$. 
The restriction gives equations (\ref{int4}), where
$N$ should be substituted by $N_0$, with $x_i$ 
($i=1, \ldots , N_0$) being the
coordinate of the $i$th pair moving as a whole thing 
with the fixed distance
between the two particles. 

In this paper we provide evidence of integrability of the deformed
RS system (\ref{int4}). To wit, we obtain the complete set 
of independent integrals
of motion in the explicit form. Our method is based on the fact 
(which is proved in the paper) that the subspace
${\cal P}$ is preserved not only by the flows with the Hamiltonian
${\sf H}_1^-$ but also by all higher Hamiltonian flows with the 
Hamiltonians ${\sf H}_k^-$. (However, the flows with the 
Hamiltonians ${\sf H}_k^+$ do not preserve the space ${\cal P}$.)
This gives the possibility to obtain the integrals of motion of
the $N_0$-particle deformed RS system by restriction of the known
integrals of motion for the $2N_0$-particle RS system to the subspace
${\cal P}$ of pairs, and this is what we do in the present paper. 

The main result of this paper is the following explicit expressions
for integrals of motion of the system (\ref{int4}) 
(with $g=\sigma (2\eta )$):
\beq\label{int6}
\begin{array}{l}
\displaystyle{
J_n=\frac{1}{2}\sum_{m=0}^{[n/2]}\frac{\sigma (n\eta )
\sigma ^{2m-n}(\eta )}{m! \, (n\! -\! 2m)!}\sum_{[i_1 \ldots i_{n-m}]}^N
\! \dot x_{i_{m+1}}\ldots \dot x_{i_{n-m}}
\!\!\!\! \prod_{\alpha , \beta =m+1 \atop \alpha <\beta}^{n-m}
\! V(x_{i_{\alpha}i_{\beta}})}
\\ \\
\displaystyle{\phantom{aaaaaaaaaaaaaaaa}
\times \left [ \prod_{\gamma =1}^m
\prod_{\ell \neq i_1, \ldots , i_{n-m}}^N\!\!\!\!
U^{+}(x_{i_{\gamma}\ell})+
\prod_{\gamma =1}^m
\prod_{\ell \neq i_1, \ldots , i_{n-m}}^N\!\!\!\!
U^{-}(x_{i_{\gamma}\ell})
\right ],
}
\end{array}
\eeq
where
$$
V(x_{ij})=\frac{\sigma^2(x_{ij})}{\sigma (x_{ij}
\! +\! \eta )\sigma (x_{ij}\! -\! \eta )}
$$
and $U^{\pm}(x_{ij})$ is given in (\ref{int5}).
In (\ref{int6}) $n=1, \ldots , N$ and $\displaystyle{
\sum_{[i_1 \ldots i_{n-m}]}^N}$ means summation over all distinct
indexes $i_1, \ldots , i_{n-m}$ from $1$ to $N$; $[n/2]$ is the 
integer part of $n/2$. At $m=0$, the product $\displaystyle{
\prod_{\gamma =1}^0}$ in the second line of (\ref{int6}) 
should be put equal to $1$. Similarly, at
$2m=n$ the product
$\dot x_{i_{m+1}}\ldots \dot x_{i_{n-m}}$ should also 
be put equal to $1$. Here are some examples for small values of $n$:
\beq\label{int7}
\begin{array}{l}
\displaystyle{
J_1=\sum_{i=1} \dot x_i,}
\\ \\
\displaystyle{
J_2=\frac{\sigma (2\eta )}{2\sigma^2(\eta )}\left [
\sum_{i\neq j} \dot x_i \dot x_j V(x_{ij})+\sigma^2(\eta )
\sum_i \Bigl (\prod_{\ell\neq i}U^+(x_{i\ell})+
\prod_{\ell\neq i}U^-(x_{i\ell})\Bigr )\right ],}
\\ \\
\!\!\! \begin{array}{l}
\displaystyle{
J_3=\frac{\sigma (3\eta )}{6\sigma^3(\eta )}\left [
\sum_{i\neq j,k, \, j\neq k} \dot x_i \dot x_j \dot x_k V(x_{ij})V(x_{ik})V(x_{jk})\right. }
\\ 
\displaystyle{\phantom{aaaaaaaaaaaaaaaaaaaa}
\left. +\,\, 3\sigma^2(\eta )
\sum_{i\neq j} \dot x_j \Bigl (\prod_{\ell\neq i,j}U^+(x_{i\ell})+
\prod_{\ell\neq i,j}U^-(x_{i\ell})\Bigr )\right ].}
\end{array}
\end{array}
\eeq
Note that the $m=0$ term in (\ref{int6}) is the $n$th 
integral of motion of the RS system (\ref{int2}).

We also find the generating function of the integrals of motion:
\beq\label{g9a}
R(z, \lambda )=\det_{1\leq i,j\leq N}
\Bigl (z\delta_{ij}-\dot x_i \phi(x_{ij}\! -\! \eta , \lambda )-
\sigma (2\eta )z^{-1} U_i^- \phi(x_{ij}\! -\! 2\eta , \lambda )\Bigr ),
\eeq
where
\beq\label{g9b}
\phi (x, \lambda ):=
\frac{\sigma (x+\lambda )}{\sigma (\lambda )\sigma (x)}.
\eeq
The equation $R(z, \lambda )=0$ defines the spectral curve which 
is an integral of motion.

The organization of the paper is as follows. In Section 2 we remind
the main facts about the elliptic RS model. In Section 3 we show,
reproducing the result of \cite{KZ22}, that the dynamics
of the deformed RS system is the 
${\sf H}_1^{-}$-flow of the RS system restricted to the space of pairs.
The core of the paper is Section 4, where we prove that the space of
pairs is invariant under all higher ${\sf H}^{-}_k$-flows and find
integrals of motion of the deformed RS system in the explicit form.
The generating function of the integrals of motion is found in
Section 5.
In Section 6 we make concluding remarks and list some open problems. 
There are also two appendices. In Appendix A the definition and main 
properties of the Weierstrass functions are presented. In Appendix B 
we prove an identity for elliptic functions which is the key identity
for the proof of Theorem \ref{theorem:inv} in Section 4. 

This paper has grown up from our joint works \cite{KZ22,KZ22a}
with Igor Krichever. Soon after the present work was started, my older
friend and co-author 
Igor Krichever passed away. He worked till the last his days, and we had
several illuminating conversations. With sorrow and gratefulness, I dedicate
this paper to his memory.

\section{The RS system}

Here we collect the main facts on the elliptic RS system 
following the paper \cite{Ruij87}.

The $N$-particle elliptic RS system
is a completely integrable model.
The canonical Poisson brackets between coordinates and momenta are
$\{x_i, p_j\}=\delta_{ij}$.
The integrals of motion in involution have the form
\beq\label{rs1}
{\sf I}_n= \sum_{{\cal I}\subset \{1, \ldots , N\}, \, |{\cal I}|=n}
\exp \Bigl (\sum_{i\in {\cal I}}p_i\Bigr ) 
\prod_{i\in {\cal I}, j\notin {\cal I}}\frac{\sigma
(x_{ij}+\eta )}{\sigma (x_{ij})}, \quad n=1, \ldots , N.
\eeq
It is natural to put ${\sf I}_0=1$.
Important particular cases of (\ref{rs1}) are
\beq\label{rs2}
{\sf I}_1=\sum_{i=1}^N e^{p_i}\prod_{j\neq i} \frac{\sigma
(x_{ij}+\eta )}{\sigma (x_{ij})}
\eeq
which is the Hamiltonian ${\sf H}_1$ of the chiral RS model and
\beq\label{rs2a}
{\sf I}_N=\exp \Bigl (\sum_{i=1}^{N}p_i\Bigr ).
\eeq
Comparing to the paper \cite{Ruij87}, our formulas differ by the canonical transformation
$$
e^{p_i}\to e^{p_i}\prod_{j\neq i}\frac{\sigma^{1/2}
(x_{ij}+\eta )}{\sigma^{1/2} (x_{ij}-\eta )}, \quad x_i\to x_i,
$$
which allows one to eliminate square roots in the formulas from 
\cite{Ruij87}.

Let us denote the time variable of the Hamiltonian 
flow with the Hamiltonian ${\sf H}_1=I_1$
by $t_1$.
The velocities of the particles are
\beq\label{rs4}
\stackrel{*}x_i =\frac{\p {\sf H}_1}{\p p_i}=
e^{p_i}\prod_{j\neq i} \frac{\sigma
(x_{ij}+\eta )}{\sigma (x_{ij})},
\eeq
where star means the $t_1$-derivative.
Note that in terms of velocities the integrals of motion (\ref{rs1})
read:
\beq\label{rs6}
{\sf I}_n=\frac{1}{n!}
\sum_{[i_1 \ldots i_n]}^N \!\! 
\stackrel{*}x_{i_1}\ldots \stackrel{*}x_{i_n}
\!\!\! \prod_{\alpha , \beta =1 \atop \alpha <\beta}^{n}
\! \frac{\sigma^2(x_{i_{\alpha}i_{\beta}})}{\sigma (x_{i_{\alpha}i_{\beta}}
\! +\! \eta )\sigma (x_{i_{\alpha}i_{\beta}}\! -\! \eta )}.
\eeq
Here $\displaystyle{
\sum_{[i_1 \ldots i_{n}]}^N}$ means summation over all distinct
indexes $i_1, \ldots , i_{n}$ from $1$ to $N$.
It is not difficult to verify that 
the Hamiltonian equations $\stackrel{*}p_i=-\p {\sf H}_1/ \p x_i$ 
are equivalent to the following
equations of motion:
\beq\label{rs7}
\stackrel{**}x_i +\sum_{k\neq i}^N
\stackrel{*}x_i\stackrel{*}x_k \Bigl (
\zeta (x_{ik}+\eta )+\zeta (x_{ik}-\eta )-2\zeta (x_{ik})\Bigr )=0
\eeq
which are equations (\ref{int2}). 

One can also introduce integrals of motion ${\sf I}_{-n}$ as
\beq\label{rs8}
{\sf I}_{-n}={\sf I}_{N}^{-1}{\sf I}_{N-n}=
\sum_{{\cal I}\subset \{1, \ldots , N\}, \, |{\cal I}|=n}
\exp \Bigl (-\sum_{i\in {\cal I}}p_i\Bigr ) \prod_{i\in {\cal I}, 
j\notin {\cal I}}\frac{\sigma
(x_{ij}-\eta )}{\sigma (x_{ij})}.
\eeq
In particular,
\beq\label{rs9}
{\sf I}_{-1}= \sum_{i=1}^N e^{-p_i}\prod_{j\neq i} \frac{\sigma
(x_{ij}-\eta )}{\sigma (x_{ij})}.
\eeq
It can be easily verified that equations of motion
in the time $\bar t_1$ corresponding to the Hamiltonian 
$\bar {\sf H}_1=\sigma^2(\eta ){\sf I}_{-1}$
are the same
as (\ref{int2}).

Let us introduce the renormalized integrals of motion:
\beq\label{rs10}
{\sf J}_n =\frac{\sigma (|n|\eta )}{\sigma^n(\eta )}\, {\sf I}_n,
\quad n=\pm 1, \ldots , \pm N.
\eeq
In the paper \cite{PZ21} it was shown that 
the higher Hamiltonians of the RS model 
can be obtained from the equation of the spectral curve
\beq\label{rs10a}
z^N +\sum_{n=1}^N \phi_n(\lambda )\, {\sf J}_n \, z^{N-n}=0, \quad 
\phi_n(\lambda )=\frac{\sigma (\lambda -n\eta )}{\sigma (\lambda )
\sigma (n\eta )}
\eeq
as
\beq\label{rs10b}
{\sf H}_n=\res_{z=\infty}\Bigl (z^{n-1}\lambda (z)\Bigr ).
\eeq
In general, they are expressed as
\beq\label{rs11}
\begin{array}{l}
{\sf H}_n={\sf J}_n +Q_n({\sf J}_1, \ldots , {\sf J}_{n-1}),
\\ \\
\bar {\sf H}_n={\sf J}_{-n} +Q_n({\sf J}_{-1}, \ldots , {\sf J}_{-n+1})
\end{array}
\eeq
for $n\in \NN$, 
where $Q_n$ are some homogeneous polynomials of homogeneity $n$ (with
degree of ${\sf J}_k$ being put equal to $k$). For example:
\beq\label{rs12}
\begin{array}{l}
{\sf H}_1={\sf J}_1,
\\ \\
{\sf H}_2={\sf J}_2 -\zeta (\eta ){\sf J}_1^2,
\\ \\
{\sf H}_3={\sf J}_3 -(\zeta (\eta )+\zeta (2\eta )){\sf J}_1{\sf J}_2
+\Bigl (\frac{3}{2}\, \zeta^2(\eta )-\frac{1}{2}\, \wp (\eta )\Bigr )
{\sf J}_1^3
\end{array}
\eeq
(see \cite{PZ21}). We also introduce the Hamiltonians
\beq\label{rs13}
{\sf H}_n^{\pm}={\sf H}_n \pm \bar {\sf H}_n.
\eeq
On the Toda lattice side, the RS dynamics corresponds to the dynamics
of poles of elliptic solutions and 
the Hamiltonians ${\sf H}_n^{\pm}$ generate
the flows $\p_{t_n}\pm \p_{\bar t_n}$, where $t_n$, $\bar t_n$ are 
canonical higher times of the Toda lattice hierarchy. 

\section{The deformed RS model as a dynamical system for pairs 
of the RS particles}

\begin{figure}[t]
\centering{\includegraphics[scale=2.0]{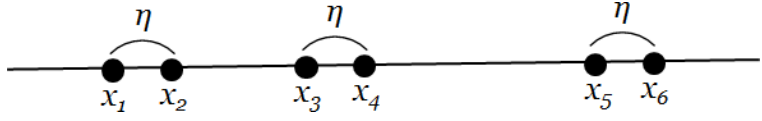}}
\vspace{1cm}
\caption{Pairs of RS particles ($N=6$, $N_0=3$).}
\label{figure:pairs}
\end{figure}

In this section we reproduce the result of \cite{KZ22} and show 
that the restriction
of the RS
dynamics of $N=2N_0$ particles to the subspace ${\cal P}$ in which the
particles stick together in $N_0$ pairs such that
\beq\label{rest1}
x_{2i}-x_{2i-1}=\eta , \qquad i=1, \ldots , N_0
\eeq
leads to the equations of motion of the deformed RS system for 
coordinates of the pairs. 
It is natural to introduce the variables 
\beq\label{rest2}
X_i =x_{2i-1}, \qquad i=1, \ldots , N_0
\eeq
which are coordinates of the pairs. 
It was proved in \cite{KZ22} that 
such structure is preserved
by the ${\sf H}_1^-$-flow $\p_t=\p_{t_1}-\p_{\bar t_1}$
but is destroyed 
by the ${\sf H}_1^+$-flow 
$\p_{t_1}+\p_{\bar t_1}$. Therefore, to define the dynamical system
we should fix $T_1^+=\frac{1}{2}\, (t_1 +\bar t_1)$ to be $0$, i.e.
put $\bar t_1 =-t_1$, and consider the evolution with respect to
the time $t=T_1^-= \frac{1}{2}\, (t_1 -\bar t_1)$.

For the velocities $\dot x_i =\p {\sf H}_1^-/\p p_i$ we have:
\beq\label{rest3a}
\dot x_{2i-1}=e^{p_{2i-1}}\! \prod_{j=1, \neq 2i-1}^{2N_0} \! 
\frac{\sigma
(x_{2i-1, j}+\eta )}{\sigma (x_{2i-1, j})}+
\sigma^2(\eta )e^{-p_{2i-1}}\! \prod_{j=1, \neq 2i-1}^{2N_0} \! 
\frac{\sigma
(x_{2i-1, j}-\eta )}{\sigma (x_{2i-1, j})},
\eeq
\beq\label{rest3b}
\dot x_{2i}=e^{p_{2i}}\! \prod_{j=1, \neq 2i}^{2N_0} \! \frac{\sigma
(x_{2i, j}+\eta )}{\sigma (x_{2i, j})}\, +\, 
\sigma^2(\eta )e^{-p_{2i}}\! \prod_{j=1, \neq 2i}^{2N_0} \! \frac{\sigma
(x_{2i, j}-\eta )}{\sigma (x_{2i, j})}.
\eeq
Under the constraint (\ref{rest1}) the first term in the right hand side
of (\ref{rest3a}) vanishes. The second term in the right hand side of
(\ref{rest3b}) also vanishes. Then
in terms of coordinates $X_i$ of the pairs equations (\ref{rest3a}),
(\ref{rest3b}) read:
\beq\label{rest3}
\begin{array}{l}
\displaystyle{
\dot x_{2i-1}=\sigma (\eta )\sigma (2\eta )e^{-p_{2i-1}}
\! \prod_{j=1, \neq i}^{N_0} \! 
\frac{\sigma (X_{ij}-2\eta )}{\sigma (X_{ij})}},
\\ \\
\displaystyle{
\dot x_{2i}=\frac{\sigma (2\eta )}{\sigma (\eta )}
\, e^{p_{2i}}
\! \prod_{j=1, \neq i}^{N_0} \! 
\frac{\sigma (X_{ij}+2\eta )}{\sigma (X_{ij})}}.
\end{array}
\eeq
From (\ref{rest3}) it is clear that if we set
\beq\label{rest4}
p_{2i-1}=\alpha_i +P_i, \quad p_{2i}=\alpha_i -P_i, \quad i=1, \ldots , N_0,
\eeq
where
\beq\label{rest5}
\alpha_i= \log \sigma (\eta )+\frac{1}{2}\sum_{j\neq i}^{N_0}
\log \frac{\sigma (X_{ij}-2\eta )}{\sigma (X_{ij}+2\eta )}
\eeq
and $P_i$ are arbitrary, then we have $\dot x_{2i-1}=
\dot x_{2i}$ for any $i$, so the distance between
the particles in each pair is preserved by the dynamics.
Under the ${\sf H}_1^{-}$-flow
each pair moves as a whole thing. 
Equations (\ref{rest3}) are then equivalent to the single equation
\beq\label{rest6}
\dot X_i=\sigma (2\eta )e^{-P_i}\prod_{j\neq i}^{N_0}
\frac{(\sigma (X_{ij}-2\eta )
\sigma (X_{ij}+2\eta ))^{1/2}}{\sigma (X_{ij})}.
\eeq

We have passed from the initial $4N_0$-dimensional phase space 
${\cal F}$ with coordinates $(\{x_i\}_N, \{p_i\}_N)$ to the 
$2N_0$-dimensional subspace
${\cal P}\subset {\cal F}$ of pairs defined
by the constraints
\beq\label{rest6a}
\left \{ \begin{array}{l}
x_{2i}-x_{2i-1}=\eta , \quad x_{2i-1}=X_i,
\\ \\
\displaystyle{p_{2i-1}+p_{2i}=
2\log \sigma (\eta )+\sum_{j\neq i}
\log \frac{\sigma (X_{ij}-2\eta )}{\sigma (X_{ij}+2\eta )}}.
\end{array} \right.
\eeq
The coordinates in ${\cal P}$ are 
$(\{X_i\}_{N_0}, \{P_i\}_{N_0})$. 

\begin{proposition}
\label{proposition:lagrangian}
The space ${\cal P}\subset {\cal F}$ defined by (\ref{rest6a})
is Lagrangian.
\end{proposition}

\noindent
{\it Proof.} We should prove that the restriction of the canonical 2-form
$\displaystyle{\Omega =\sum_{i=1}^{2N} dp_i \wedge dx_i}$ to the 
half-dimensional subspace
${\cal P}$ is identically zero. This is a simple calculation with the 
help of equations (\ref{rest4}), (\ref{rest5}) and (\ref{rest6a}).
\square

\begin{theorem}
The subspace ${\cal P}$ is preserved by the Hamiltonian flow
with the Hamiltonian
${\sf H}_1^{-}={\sf H}_1-\bar {\sf H}_1$ and 
equations of motion of the deformed RS model (\ref{int4}) are 
obtained as the restriction of this flow to the subspace ${\cal P}$.
\end{theorem}

\noindent
{\it Proof.}
Restricting the second set of the Hamiltonian equations, 
$\dot p_i=-\p {\sf H}_1^-/\p x_i$, to the subspace ${\cal P}$, we have:
\beq\label{rest7}
\begin{array}{lll}
\dot p_{2i-1} & \!\! =\!\! & 
\displaystyle{
\sigma (\eta )\sigma (2\eta )e^{-\alpha_i -P_i}\!\! \prod_{k=1, \neq i}^{N_0}
\!\!
\frac{\sigma (X_{ik}-2\eta )}{\sigma (X_{ik})}\left [
\sum_{j=1, \neq i}^{N_0}\Bigl 
(\zeta (X_{ij}\! -\! 2\eta )\! -\! \zeta (X_{ij})\Bigr )
\! +\! \zeta (\eta ) \! -\! \zeta (2\eta )\right ]}
\\ &&\\
&& +\, \displaystyle{
\sigma (\eta )\sigma (2\eta )\sum_{l=1, \neq i}^{N_0}
e^{-\alpha_l -P_l}\! \prod_{k=1, \neq l}^{N_0}
\frac{\sigma (X_{lk}-2\eta )}{\sigma (X_{lk})}
\Bigl (\zeta (X_{il}+\eta )-\zeta (X_{il})\Bigr )}
\\ &&\\
&& -\, \displaystyle{
\frac{\sigma (2\eta )}{\sigma (\eta )}\sum_{l=1}^{N_0}
e^{\alpha_l -P_l}\! \prod_{k=1, \neq l}^n
\frac{\sigma (X_{lk}+2\eta )}{\sigma (X_{lk})}
\Bigl (\zeta (X_{il}-2\eta )-\zeta (X_{il}-\eta )\Bigr )}
\\ && \\
&& +\, \displaystyle{
\sigma^{-1}(\eta )e^{\alpha_i +P_i}\prod_{k=1, \neq i}^{N_0}
\frac{\sigma (X_{ik}+\eta )}{\sigma (X_{ik})-\eta )}-
\sigma (\eta )e^{-\alpha_i +P_i}\prod_{k=1, \neq i}^{N_0}
\frac{\sigma (X_{ik}-\eta )}{\sigma (X_{ik})+\eta )}}.
\end{array}
\eeq
Taking the time derivative of (\ref{rest6}), we obtain:
\beq\label{rest7a}
\begin{array}{l}
\displaystyle{
\ddot X_i=-\sigma (2\eta )\dot P_i e^{-P_i}
\prod_{j\neq i}^{N_0}
\frac{(\sigma (X_{ij}-2\eta )
\sigma (X_{ij}+2\eta ))^{1/2}}{\sigma (X_{ij})}}
\\ \\
\displaystyle{
+\frac{1}{2}\sum_{j\neq i}^{N_0}\dot X_i (\dot X_i-\dot X_j)
\Bigl (\zeta (X_{ij}-2\eta )+\zeta (X_{ij}+2\eta )-2\zeta (X_{ij})\Bigr ),}
\end{array}
\eeq
where we should substitute $\dot P_i=-\dot \alpha_i
+\dot p_{2i-1}$ from (\ref{rest7}) taking into account
(\ref{rest6}):
$$
\dot P_i =-\dot \alpha_i +\dot X_i \left [\sum_{j\neq i}^{N_0}
\Bigl (\zeta (X_{ij}-2\eta )-\zeta (X_{ij})\Bigr ) +\zeta (\eta )-
\zeta (2\eta )\right ]
$$
$$
+\sum_{l\neq i}^{N_0}
\dot X_l \Bigl (\zeta (X_{il}+\eta )-\zeta (X_{il})\Bigr )
-\sum_{l=1}^{N_0}\dot X_l \Bigl (\zeta (X_{il}-2\eta )-
\zeta (X_{il}-\eta )\Bigr )
$$
$$
+e^{P_i}\prod_{k\neq i}^{N_0}
\frac{\sigma^{1/2}(X_{ik}-2\eta )
\sigma (X_{ik}+\eta )}{\sigma^{1/2}(X_{ik}+2\eta )\sigma (X_{ik}-\eta )}
-e^{P_i}\prod_{k\neq i}^{N_0}\frac{\sigma^{1/2}(X_{ik}+2\eta )
\sigma (X_{ik}-\eta )}{\sigma^{1/2}(X_{ik}-2\eta )\sigma (X_{ik}+\eta )}.
$$
Plugging here $\dot \alpha_i$ from (\ref{rest5}) and substituting into
(\ref{rest7a}),
we finally obtain:
\beq\label{rest8}
\ddot X_i=-\sum_{j\neq i}^{N_0}\dot X_i \dot X_j \Bigl (
\zeta (X_{ij}+\eta )+\zeta (X_{ij}-\eta )-2\zeta (X_{ij})\Bigr )
+\sigma (2\eta )\Bigl (U_i^+ - U_i^-\Bigr ),
\eeq
where
\beq\label{rest9}
U_i^{\pm}=\prod_{j\neq i}^{N_0}\frac{\sigma (X_{ij}\pm 2\eta )
\sigma (X_{ij}\mp \eta )}{\sigma
(X_{ij}\pm \eta )\sigma (X_{ij})}.
\eeq
These are equations (\ref{int1}), (\ref{int2}) of the deformed 
RS system (at $g=\sigma (2\eta )$, $N=N_0$). 
\square

\section{Integrals of motion}

In this section we are going to 
prove that the subspace ${\cal P}$ is invariant not only 
with respect to the ${\sf H}_1^{-}$-flow but also with respect to
all higher ${\sf H}_k^{-}$-flows. This gives the possibility to
obtain integrals of motion $J_n$ of the deformed RS model by 
restriction of the RS integrals of motion ${\sf J}_n$, ${\sf J}_{-n}$
to the subspace ${\cal P}$. 
We denote the restriction of ${\sf J}_k$ by $J_k$:
\beq\label{im0}
J_k ((\{X_i\}_{N_0}, \{P_i\}_{N_0})=
{\sf J}_k (\{x_{\ell}\}_N, \{p_{\ell}\}_N)\Bigr |_{{\cal P}}, 
\quad k\in \ZZ .
\eeq
The notation 
${\sf J}_k (\{x_{\ell}\}_N, \{p_{\ell}\}_N)\Bigr |_{{\cal P}}$ means that
the variables $x_{\ell}$, $p_{\ell}$ are constrained by the relations
(\ref{rest6a}), i.e.
$$
x_{2i-1}=X_i, \quad x_{2i}=X_i +\eta ,
$$
$$
p_{2i-1}=\alpha_i (\{X_j\}_{N_0}) +P_i, \quad p_{2i}=
\alpha_i (\{X_j\}_{N_0}) -P_i,
$$
where $\alpha_i$ is given by (\ref{rest5}).
Note that $J_k$ can be regarded as a function of $\{X_j\}_{N_0}$ and
$\{\dot X_j\}_{N_0}$ by virtue of equation (\ref{rest6}) and
$$
\frac{\p J_k}{\p P_i}=-\dot X_i \frac{\p J_k}{\p \dot X_i}.
$$
The similar notation will be used for 
the restriction of the Hamiltonians:
\beq\label{im0a}
\begin{array}{l}
H_k (\{X_i\}_{N_0}, \{P_i\}_{N_0}) =
{\sf H}_k (\{x_{\ell}\}_N, \{p_{\ell}\}_N)\Bigr |_{{\cal P}},
\\ \\
\bar H_k (\{X_i\}, \{P_i\}) =\bar {\sf H}_k 
(\{x_{\ell}\}_N, \{p_{\ell}\}_N)\Bigr |_{{\cal P}}.
\end{array}
\eeq

\begin{theorem}\label{theorem:inv}
The space ${\cal P}$ of pairs defined by (\ref{rest6a}) 
is invariant with respect to the 
Hamiltonian flows $\p_{t_k}-\p_{\bar t_k}$ with the Hamiltonians
${\sf H}_k^{-}$ for all $k\geq 1$.
\end{theorem}

\noindent
The rest of this section is devoted to the proof of 
Theorem \ref{theorem:inv}.
The explicit expressions for integrals of motion of the deformed 
RS system will follow from the proof. 

To prove that the first constraint, $x_{2i-1}-x_{2i}=\eta$, is preserved,
we should show that $(\p_{t_k}-\p_{\bar t_k})x_{2i-1}=
(\p_{t_k}-\p_{\bar t_k})x_{2i}$ for all
$i=1, \ldots , N_0$, i.e. that
\beq\label{im1}
\frac{\p {\sf H}_k}{\p p_{2i-1}}-\frac{\p \bar {\sf H}_k}{\p p_{2i-1}}=
\frac{\p {\sf H}_k}{\p p_{2i}}-\frac{\p \bar {\sf H}_k}{\p p_{2i}}
\eeq
if the coordinates and momenta are restricted to the space ${\cal P}$.
Note that equations (\ref{rest4}) imply that 
$\p_{p_{2i-1}}-\p_{p_{2i}}=\p_{P_i}$, so
(\ref{im1}) is equivalent to
\beq\label{im2}
\frac{\p H_k}{\p P_i}=\frac{\p \bar H_k}{\p P_i}.
\eeq
From (\ref{rs11}) it follows that it is enough to prove that
$J_n =J_{-n}$. 

Let ${\cal N}$ be the set
${\cal N}=\{1, \ldots , N_0\}$. Separating the summation in (\ref{rs1})
over odd and even indexes (with $m$ odd indexes and $n-m$ even
ones), we can write, for $0<n\leq N_0$:
\beq\label{im3a}
J_{n}=\sum_{m=0}^n J_{n,m},
\eeq
where
\beq\label{im3}
\begin{array}{l}
\displaystyle{
J_{n,m}=\frac{\sigma (n\eta )}{\sigma^n(\eta )}\, 
\sum_{{\cal I}, {\cal J}\subseteq {\cal N}
\atop {|{\cal I}|=m \atop |{\cal J}|=n-m}}\!
\Bigl (\prod_{i\in {\cal I}}e^{p_{2i-1}}\Bigr )
\Bigl (\prod_{j\in {\cal J}}e^{p_{2j}}\Bigr )}
\\ \\
\displaystyle{
\times \prod_{\ell \in {\cal N}\setminus {\cal I}}\,\, 
\prod_{i\in {\cal I}}
\frac{\sigma (X_{i\ell}+\eta )}{\sigma (X_{i\ell})}\, 
\prod_{\ell \in {\cal N}\setminus {\cal I}}\, \prod_{j\in {\cal J}}
\frac{\sigma (X_{j\ell}+2\eta )}{\sigma (X_{i\ell}+\eta )}}
\\ \\
\displaystyle{
\times \prod_{\ell \in {\cal N}\setminus {\cal J}}\,\, 
\prod_{i\in {\cal I}}
\frac{\sigma (X_{i\ell})}{\sigma (X_{i\ell}-\eta )}\,
\prod_{\ell \in {\cal N}\setminus {\cal J}}\, \prod_{j\in {\cal J}}
\frac{\sigma (X_{j\ell}+\eta )}{\sigma (X_{j\ell})}}.
\end{array}
\eeq
Obviously, this is zero unless 
${\cal I}\cap ({\cal N}\setminus {\cal J})=\emptyset$, i.e. 
the set ${\cal I}$ should be contained in ${\cal J}$, 
${\cal I}\subseteq {\cal J}$. Since $|{\cal I}|=m$, $|{\cal J}|=n-m$,
this is possible only if $m\leq [n/2]$, otherwise $J_{n,m}$ vanishes. 
Using (\ref{rest4}), (\ref{rest5}), (\ref{rest6}), we then have:
$$
\Bigl (\prod_{i\in {\cal I}}e^{p_{2i-1}}\Bigr )
\Bigl (\prod_{j\in {\cal J}}e^{p_{2j}}\Bigr )
$$
$$
=\frac{\sigma^n(\eta )}{\sigma^{n-2m}(2\eta )}
\Bigl (\prod_{i\in {\cal I}}\, \prod_{\ell \in {\cal N}\setminus \{i\}}
\frac{\sigma (X_{i\ell}-2\eta )}{\sigma (X_{i\ell}+2\eta )}\Bigr )\,
\Bigl (\prod_{j\in {\cal J}\setminus {\cal I}}\, 
\prod_{\ell \in {\cal N}\setminus \{j\}}
\frac{\sigma (X_{j\ell})}{\sigma (X_{j\ell}+2\eta )}\Bigr )
\prod_{j\in {\cal J}\setminus {\cal I}}\dot X_j.
$$
The expression for $J_{-n,m}$ is similar but in this case $m$
is the number of even indexes rather than odd and $\eta$ in all
factors in the products should be
replaced by $-\eta$. After plugging this into (\ref{im3}) 
and cancellations, we obtain:
\beq\label{im4}
J_{\pm n, m}=\frac{\sigma (n\eta )}{\sigma ^{n-2m}(\eta )}
\sum_{{\cal J}\atop |{\cal J}|=n-m} \,
\sum_{{\cal I}\subseteq {\cal J}\atop |{\cal I}|=m}
\Bigl (\prod_{j\in {\cal J}\setminus {\cal I}}\dot X_j\Bigr )
\Bigl (\prod_{i,j\in {\cal J}\setminus {\cal I}\atop i<j}
V(X_{ij})\Bigr ) \Bigl (\prod_{i\in {\cal I}}\prod_{\ell \in {\cal N}
\setminus {\cal J}}U^{\pm}(X_{i\ell} )\Bigr ),
\eeq
where
\beq\label{im5}
V(X_{ij})=\frac{\sigma^2(X_{ij})}{\sigma (X_{ij}+\eta )\, 
\sigma (X_{ij}-\eta )},
\eeq
\beq\label{im6}
U^{\pm}(X_{ij})=\frac{\sigma (X_{ij}\pm 2\eta )\,
\sigma (X_{ij}\mp \eta )}{\sigma (X_{ij}\pm \eta )\, \sigma (X_{ij})}.
\eeq
Passing from summation over the subsets ${\cal J}\subset {\cal N}$ 
and ${\cal I}\subseteq {\cal J}$ to the summation over 
subsets ${\cal I}$ and
${\cal I}'$ such that ${\cal I}\cap {\cal I}'=\emptyset$
(${\cal I}'={\cal J}\setminus {\cal I}$), we can write
the r.h.s. of (\ref{im4}) in the form
\beq\label{im7}
J_{\pm n, m}=\frac{\sigma (n\eta )}{\sigma ^{n-2m}(\eta )}
\sum_{{\cal I}, {\cal I}', {\cal I}\cap {\cal I}'=\emptyset
\atop |{\cal I}|=m, |{\cal I}'|=n-2m} \,
\Bigl (\prod_{j\in {\cal I}'}\dot X_j\Bigr )
\Bigl (\prod_{i,j\in {\cal I}' \atop i<j}
V(X_{ij})\Bigr ) \Bigl (\prod_{i\in {\cal I}}\, \prod_{\ell \in {\cal N}
\setminus ({\cal I}\cup {\cal I}')}\! U^{\pm}(X_{i\ell} )\Bigr ).
\eeq
The equality $J_{n,m}=J_{-n,m}$ is a consequence of the following lemma:
\begin{lemma}\label{lemma:identity}
For any ${\cal N}'\subseteq {\cal N}=\{1, \ldots , N_0\}$
it holds:
\beq\label{im8}
\sum_{{\cal I}\subset {\cal N}'}
\prod_{i\in {\cal I}}\prod_{\ell 
\in {\cal N}'\setminus {\cal I}}U^+(X_{i\ell})=
\sum_{{\cal I}\subset {\cal N}'}
\prod_{i\in {\cal I}}\prod_{\ell 
\in {\cal N}'\setminus {\cal I}}U^-(X_{i\ell}).
\eeq
\end{lemma}
The lemma is proved in Appendix B. 
Applying the lemma with ${\cal N}'={\cal N}\setminus {\cal I}'$
to (\ref{im7}), we see that $J_{n,m}=J_{-n,m}$. The formula (\ref{int6})
for the integrals of motion in the Introduction is an explicitly 
symmetrized 
version of (\ref{im7}):
$$J_n=\frac{1}{2}\sum_{m=0}^{[n/2]}(J_{n,m}+J_{-n,m}).$$
We have proved the half of the statement of Theorem \ref{theorem:inv}:
namely, that the first constraint in (\ref{rest6a}), 
$x_{2i}-x_{2i-1}=\eta$, is invariant 
under the flows $\p_{t_k}-\p_{\bar t_k}$. 

Let us prove that the second constraint in (\ref{rest6a}) 
is preserved too.
We should show that the equality in (\ref{rest6a}) 
remains true after applying
$\p_{t_n}-\p_{\bar t_n}$ to the both sides. In the l.h.s. we then have
$$
\frac{\p {\sf H}^{-}_n}{\p x_{2i-1}}+
\frac{\p {\sf H}^{-}_n}{\p x_{2i}}=
\frac{\p {\sf H}^{-}_n}{\p X_{i}}.$$
Without loss of generality we may put $i=1$ for simplicity
of the notation. Then we have to prove that 
$$
\frac{\p {\sf H}^{-}_n}{\p X_1}=
\sum_{k\neq 1} \left (\frac{\p {\sf H}^{-}_n}{\p p_1}-
\frac{\p {\sf H}^{-}_n}{\p p_{2k-1}}\right )\Bigl (
\zeta (X_{1k}+2\eta )-\zeta (X_{1k}-2\eta )\Bigr ).
$$
From (\ref{rs11}) it is clear that it is equivalent to
\beq\label{im9}
\frac{\p {\sf J}^{-}_n}{\p X_1}=
\sum_{k\neq 1} \left (\frac{\p {\sf J}^{-}_n}{\p p_1}-
\frac{\p {\sf J}^{-}_n}{\p p_{2k-1}}\right )\Bigl (
\zeta (X_{1k}+2\eta )-\zeta (X_{1k}-2\eta )\Bigr ).
\eeq
Repeating the calculation leading to (\ref{im7}) for the restriction 
of $\p {\sf J}_{\pm n}/\p p_{2k-1}$ to the subspace ${\cal P}$, we 
obtain:
\beq\label{im10}
\frac{\p {\sf J}_{n}}{\p p_{2k-1}}=
\frac{\sigma (n\eta )}{\sigma^{n-2m}(\eta )}\sum_{m=0}^{[n/2]}\,
\sum_{{\cal I}\cap {\cal I}'=\emptyset \atop |{\cal I}|=m, \,
|{\cal I}'|=n-2m}\!\! \Theta (k\in {\cal I}) X_{{\cal I}'}
U_{{\cal I}{\cal I}'}^{-},
\eeq
\beq\label{im11}
\frac{\p {\sf J}_{-n}}{\p p_{2k-1}}=
\frac{\sigma (n\eta )}{\sigma^{n-2m}(\eta )}\sum_{m=0}^{[n/2]}\,
\sum_{{\cal I}\cap {\cal I}'=\emptyset \atop |{\cal I}|=m, \,
|{\cal I}'|=n-2m}\!\! \Theta (k\in {\cal I}\cup {\cal I}') 
X_{{\cal I}'}
U_{{\cal I}{\cal I}'}^{+}.
\eeq
Here
\beq\label{im12}
X_{{\cal I}'}=\Bigl (\prod_{j\in {\cal I}'}\dot X_j\Bigr )
\prod_{j_1, j_2 \in {\cal I}'\atop j_1<j_2} V_{j_1 j_2},
\eeq
\beq\label{im13}
U_{{\cal I}{\cal I}'}^{\pm}=\prod_{i\in {\cal I}}
\prod_{j\in {\cal N}\setminus ({\cal I}\cup {\cal I}')}U^{\pm}(X_{ij})
\eeq
and $\Theta (S)$ is the function which is equal to $1$ if the 
statement $S$ is true and $0$ otherwise. 
Combining (\ref{im10}) and (\ref{im11}), we get:
\beq\label{im14}
\begin{array}{l}
\displaystyle{
\frac{\p {\sf J}_{n}^-}{\p p_{2k-1}}=
\sum_{m=0}^{[n/2]}\kappa_{nm}\left \{
\sum_{{\cal I}\cap {\cal I}'=\emptyset \atop |{\cal I}|=m, \,
|{\cal I}'|=n-2m}\!\! \Theta (k\in {\cal I}) 
X_{{\cal I}'}
\Bigl (U_{{\cal I}{\cal I}'}^{-}+U_{{\cal I}{\cal I}'}^{+}\Bigr )\right.}
\\ \\
\displaystyle{
\left. \phantom{aaaaaaaaaaaaaaa}
+\sum_{{\cal I}\cap {\cal I}'=\emptyset \atop |{\cal I}|=m, \,
|{\cal I}'|=n-2m}\!\! \Theta (k\in {\cal I}') X_{{\cal I}'}
U_{{\cal I}{\cal I}'}^{+}\right \},}
\end{array}
\eeq
where
\beq\label{im15}
\kappa_{nm}=\sigma (n\eta )\sigma ^{2m-n}(\eta ).
\eeq
A similar calculation gives
\beq\label{im16}
\frac{\p {\sf J}_{\pm n}}{\p X_1}=
\sum_{m=0}^{[n/2]}\kappa_{nm}
\!\!\! \sum_{{\cal I}\cap {\cal I}'=\emptyset \atop |{\cal I}|=m, \,
|{\cal I}'|=n-2m} \!\! X_{{\cal I}'}
U_{{\cal I}{\cal I}'}^{\mp}Z_{{\cal I}{\cal I}'}^{\pm},
\eeq
where
\beq\label{im17}
\begin{array}{c}
Z_{{\cal I}{\cal I}'}^{+}=\displaystyle{
\Theta (1\! \in \! {\cal I})\left (\sum_{\ell \in {\cal N}\setminus
({\cal I}\cup {\cal I}')}\! \Bigl (\zeta (X_{1\ell}+ \eta )-
\zeta (X_{1\ell}- \eta )\Bigr )\! +\! 
\sum_{\ell \in {\cal I}'}\Bigl (\zeta (X_{1\ell}+ \eta )-
\zeta (X_{1\ell})\Bigr )\right )}
\\ \\
+\displaystyle{
\Theta (1\! \in \! {\cal I}\cup {\cal I}')
\left (\sum_{\ell \in {\cal N}\setminus
({\cal I}\cup {\cal I}')}\! \Bigl (\zeta (X_{1\ell}+ 2\eta )-
\zeta (X_{1\ell})\Bigr )\! +\! 
\sum_{\ell \in {\cal I}'}\Bigl (\zeta (X_{1\ell}+ 2\eta )-
\zeta (X_{1\ell}+\eta )\Bigr )\right )}
\\ \\
+\, \displaystyle{
\Theta (1\! \in \! {\cal N}\setminus {\cal I})
\left (\sum_{\ell \in {\cal I}}\! \Bigl (\zeta (X_{1\ell}+ 2\eta )-
\zeta (X_{1\ell})\Bigr )\! +\! 
\sum_{\ell \in {\cal I}'}\Bigl (\zeta (X_{1\ell}+ 2\eta )-
\zeta (X_{1\ell}+\eta )\Bigr )\right )}
\\ \\
+\displaystyle{
\Theta (1\! \in \! {\cal N}\setminus ({\cal I}\cup {\cal I}'))
\left (\sum_{\ell \in {\cal I}}\! \Bigl (\zeta (X_{1\ell}+ \eta )-
\zeta (X_{1\ell}-\eta )\Bigr )\! +\! 
\sum_{\ell \in {\cal I}'}\Bigl (\zeta (X_{1\ell}+ \eta )-
\zeta (X_{1\ell})\Bigr )\right )}
\end{array}
\eeq
and $Z_{{\cal I}{\cal I}'}^{-}$ is obtained from
$Z_{{\cal I}{\cal I}'}^{+}$ by the change $\eta \to -\eta$.
This expression can be brought to a more convenient form by using
the obvious relations
$$
\Theta (1\! \in \! {\cal I}\cup {\cal I}')=
\Theta (1 \in {\cal I})+\Theta (1 \in {\cal I}'),
\quad
\Theta (1\! \in \! {\cal N}\setminus ({\cal I})=1-
\Theta (1 \in {\cal I}).
$$

The right hand sides of (\ref{im14}) and (\ref{im16}) are sums 
over $m=0, \ldots [n/2]$. Let us denote the $m$th terms of the sums
by $\displaystyle{\frac{\p {\sf J}_{n,m}^-}{\p p_{2k-1}}}$ and
$\displaystyle{\frac{\p {\sf J}_{\pm n,m}^-}{\p X_1}}$.
We are going to show that
\beq\label{im9a}
\frac{\p {\sf J}_{n,m}}{\p X_1}-\frac{\p {\sf J}_{-n,m}}{\p X_1}=
\sum_{k\neq 1} \left (\frac{\p {\sf J}^{-}_{n,m}}{\p p_1}-
\frac{\p {\sf J}^{-}_{n,m}}{\p p_{2k-1}}\right )\Bigl (
\zeta (X_{1k}+2\eta )-\zeta (X_{1k}-2\eta )\Bigr )
\eeq
from which (\ref{im9}) follows. A straightforward calculation yields:
\beq\label{im19}
\begin{array}{c}
\displaystyle{
\kappa_{nm}^{-1}\left \{\frac{\p {\sf J}_{n,m}^{-}}{\p X_1}-
\sum_{k\neq 1} \left (\frac{\p {\sf J}^{-}_{n,m}}{\p p_1}-
\frac{\p {\sf J}^{-}_{n,m}}{\p p_{2k-1}}\right )\Bigl (
\zeta (X_{1k}+2\eta )-\zeta (X_{1k}-2\eta )\Bigr )\right \}}
\\ \\
\displaystyle{
=\sum_{{\cal I}\cap {\cal I}'=\emptyset}
X_{{\cal I}'}\left \{\Theta (1 \in {\cal I})\left [
U_{{\cal I}{\cal I}'}^{-}{\sum_{\ell}}'\zeta^{-}(X_{1\ell})-
U_{{\cal I}{\cal I}'}^{+}{\sum_{\ell}}'\zeta^{+}(X_{1\ell})+
U_{{\cal I}{\cal I}'}^{+}{\sum_{{\ell}\in {\cal I}} }'\zeta^{+}(X_{1\ell})
\right. \right. }
\\ \\
\displaystyle{
+U_{{\cal I}{\cal I}'}^{+}{\sum_{{\ell}\in {\cal I}}}'\zeta^{-}(X_{1\ell})
-U_{{\cal I}{\cal I}'}^{-}{\sum_{{\ell}\in {\cal I}}}'\zeta^{-}(X_{1\ell})
-U_{{\cal I}{\cal I}'}^{-}{\sum_{{\ell}\in {\cal I}}}'\zeta^{+}(X_{1\ell})}
\\ \\
\displaystyle{
\phantom{aaa}\left.
+U_{{\cal I}{\cal I}'}^{+}{\sum_{{\ell}\in {\cal I}'}}\zeta^{+}(X_{1\ell})
-U_{{\cal I}{\cal I}'}^{-}{\sum_{{\ell}\in {\cal I}'}}\zeta^{-}(X_{1\ell})
\right ]}
\\ \\
\displaystyle{
+\,\, \Theta (1 \in {\cal I}')\left [ U_{{\cal I}{\cal I}'}^{+}{
\sum_{{\ell}\in {\cal I}}}\zeta^{-}(X_{1\ell})-
U_{{\cal I}{\cal I}'}^{-}{
\sum_{{\ell}\in {\cal I}}}\zeta^{+}(X_{1\ell})\right ]}
\\ \\
\displaystyle{\left.
+\, \sum_{k\neq 1}\Theta (k \in {\cal I})\left [ U_{{\cal I}{\cal I}'}^{-}
\zeta^{+}(X_{1k})-U_{{\cal I}{\cal I}'}^{+}
\zeta^{-}(X_{1k})\right ] \right \},}
\end{array}
\eeq
where
\beq\label{im18}
\zeta^{\pm}(X)=\zeta (X\pm 2\eta )+\zeta (X\mp \eta )-\zeta (X+\pm \eta )
-\zeta (X)
\eeq
and $\displaystyle{{\sum_{\ell}}'}$ means that 
$\ell \neq 1$.

\begin{lemma}
The following identity holds:
\beq\label{im20}
\begin{array}{c}
\displaystyle{
\sum_{{\cal I}}\Theta (1 \in {\cal I})\left [
U_{{\cal I}{\cal I}'}^{-}\sum_{\ell \in {\cal N}\setminus
({\cal I}\cup {\cal I}')\atop \ell \neq 1}\zeta^{-}(X_{1\ell})-
U_{{\cal I}{\cal I}'}^{+}\sum_{\ell \in {\cal N}\setminus
({\cal I}\cup {\cal I}')\atop \ell \neq 1}\zeta^{+}(X_{1\ell})\right.
}
\\ \\
\displaystyle{
\left. -U_{{\cal I}{\cal I}'}^{-}\sum_{\ell \in 
{\cal I}, \ell \neq 1}\zeta^{+}(X_{1\ell})+
U_{{\cal I}{\cal I}'}^{+}\sum_{\ell \in 
{\cal I}, \ell \neq 1}\zeta^{-}(X_{1\ell})\right ]
}
\\ \\
\displaystyle{
+\sum_{{\cal I}}\Theta (1 \in {\cal I}')\left [
U_{{\cal I}{\cal I}'}^{+}\sum_{\ell \in 
{\cal I}}\zeta^{-}(X_{1\ell})-
U_{{\cal I}{\cal I}'}^{-}\sum_{\ell \in 
{\cal I}}\zeta^{+}(X_{1\ell})\right ]}
\\ \\
\displaystyle{
+\sum_{{\cal I}}\left [U_{{\cal I}{\cal I}'}^{-}\sum_{\ell \in 
{\cal I}, \ell \neq 1}\zeta^{+}(X_{1\ell} )-
U_{{\cal I}{\cal I}'}^{+}\sum_{\ell \in 
{\cal I}, \ell \neq 1}\zeta^{-}(X_{1\ell} )\right ]=0},
\end{array}
\eeq
where $U_{{\cal I}{\cal I}'}^{\pm}$ and $\zeta^{\pm}(x)$ 
are defined in (\ref{im13}) and (\ref{im18}) respectively.
\end{lemma} 

\noindent
{\it Proof.} This is the $X_1$-derivative of the identity (\ref{im8})
from Lemma \ref{lemma:identity} with ${\cal N}'={\cal N}\setminus
{\cal I}'$.
\square

\noindent
Using this identity, it is easy to see that the r.h.s. of 
(\ref{im19}) is zero. Therefore, the invariance of the subspace
${\cal P}$ of pairs with respect to the flows with Hamiltonians
${\sf H}^{-}_n$ is proved. 

So far we considered the restriction of ${\cal J}_n$ with $n<N/2=N_0$.
The case $N/2 <n\leq N$ can be considered in a similar way with the
result that the restriction of ${\cal J}_n$ with $N_0 <n\leq 2N_0$
is $J_{n-2N_0}$. The proof of Theorem \ref{theorem:inv} can be extended
to this case, too.

Finally, let us comment on whether the integrals of motion
are in involution. As soon as the Hamiltonian structure of the deformed
RS system (if any) is not known, we are not able to calculate the
Poisson brackets between the integrals of motion and prove that they
are equal to zero. Our integrals of motion are functions of coordinates
and velocities rather than coordinates and momenta. However, in 
any integrable system all integrals of motion that are in involution
are conserved quantities for the flows generated by any one of them.
Each higher Hamiltonian ${\sf H}^{-}_n$ of the RS system defines
a flow $\p_{T_n^-}$ on the ``phase space'' ${\cal P}$ of the deformed
RS system. From the fact that RS integrals of motion are in involution
it follows that the restrictions $H_n$ of the RS Hamiltonians to the
space ${\cal P}$ are conserved under all $\p_{T_k^-}$-flows.
In this sense we can say that the integrals of motion $H_n$ and $J_n$
of the deformed RS system are in involution.

\section{Generating function of the integrals of motion}

It is known that the integrals of motion of the 
RS system with $2N$ particles can be
unified into a generating function which is the determinant of the
$2N \! \times \! 2N$ matrix 
$zI - L(\lambda )$, where
$I$ is the unity matrix, $z$ is the spectral parameter and
$L(\lambda )$ is the Lax matrix depending on another
spectral parameter $\lambda$. The Lax matrix has the form
\beq\label{g1}
L_{ij}(\lambda )=\p_{t_1}x_i \, \phi (x_{ij}-\eta , \lambda ),
\eeq
where the function $\phi (x, \lambda )$ is given by
\beq\label{g2}
\phi (x, \lambda )=\frac{\sigma (x+\lambda )}{\sigma (\lambda )
\sigma (x)}.
\eeq

\begin{proposition}(\cite{Ruij87})
It holds
\beq\label{g3}
\det_{1\leq i,j\leq 2N} \Bigl (z\delta_{ij}-L_{ij}(\lambda )\Bigr )=
z^{2N}+\sum_{k=1}^{2N}z^{2N-k}
\frac{\sigma (\lambda -k\eta )}{\sigma (\lambda )\sigma (k\eta )}\, 
{\sf J}_k,
\eeq
where ${\sf J}_k$ are the 
RS integrals of motion (\ref{rs10}) (see (\ref{rs10a}) with $N\to 2N$).
\end{proposition}

\noindent
Proof of this proposition is based on the formula for the 
determinant of the elliptic Cauchy matrix:
\beq\label{g4}
\det_{1\leq i,j\leq n}\phi (y_i-x_j)=
\frac{\sigma 
\Bigl (\lambda +\sum\limits_{k=1}^n (y_k-x_k)\Bigr )}{\sigma (\lambda )}
\, \frac{\prod\limits_{k<l}
\sigma (y_k-y_l)\sigma (x_l-x_k)}{\prod\limits_{k,l}\sigma (y_k-x_l)}.
\eeq

In this section we are going to construct the generating function
for the integrals of motion (\ref{int6}). The idea is to restrict
the Lax matrix (\ref{g1}) to the subspace ${\cal P}$. 
However, the direct restriction is not possible because some matrix
elements become infinite. Nevertheless, we shall see that the determinant
(\ref{g3}) is finite. 

To regularize the Lax matrix, we put
\beq\label{g5}
x_{2i}-x_{2i-1}=\eta +\varepsilon
\eeq
and tend $\varepsilon \to 0$ at the end. At 
$\varepsilon =0$ we have $\p_{t_1}x_{2i}=\dot X_i$
and $\p_{t_1}x_{2i-1}=0$. To proceed, 
we need to find $\p_{t_1}x_{2i-1}$ up to 
the first non-vanishing order in $\varepsilon$. A simple calculation
shows that
\beq\label{g6}
\p_{t_1}x_{2i-1}=\varepsilon \sigma (2\eta )
\dot X_i^{-1}U_i^- +O(\varepsilon^2),
\eeq
where $U_i^-$ is given by (\ref{rest9}) (with $N_0\to N$). 
The further calculation of matrix elements of the Lax matrix 
is straightforward:
\beq\label{g7}
\begin{array}{l}
L_{2i-1, 2j-1}:=L_{ij}^{({\rm oo})}=\varepsilon \sigma (2\eta )
\dot X_i^{-1}U_i^-\phi (X_{ij}-\eta , \lambda ) +O(\varepsilon^2),
\\ \\
L_{2i-1, 2j}:=L_{ij}^{({\rm oe})}=\varepsilon \sigma (2\eta )
\dot X_i^{-1}U_i^-\phi (X_{ij}-2\eta , \lambda ) +O(\varepsilon^2),
\\ \\
L_{2i, 2j-1}:=L_{ij}^{({\rm eo})}=
\dot X_i \phi (X_{ij}+\varepsilon , \lambda )+
\delta_{ij}O(1)+O(\varepsilon ),
\\ \\
L_{2i, 2j}:=L_{ij}^{({\rm ee})}=
\dot X_i \phi (X_{ij}-\eta , \lambda )+O(\varepsilon ).
\end{array}
\eeq
After re-numeration of rows and columns,
the Lax matrix can be represented as a $2\! \times \! 2$ block matrix:
\beq\label{g8}
L=\left ( \begin{array}{ll}
L_{ij}^{({\rm oo})} & L_{ij}^{({\rm oe})}
\\ \\
L_{ij}^{({\rm eo})} & L_{ij}^{({\rm ee})}
\end{array}\right ), \quad i,j =1, \ldots , N.
\eeq
We see that $L_{ii}^{({\rm eo})}$ is singular as $\varepsilon \to 0$
since $\phi (\varepsilon , \lambda )=\varepsilon^{-1}+O(1)$.
Using the formula for determinant of a block matrix,
we have:
$$
\det (zI -L)=\det \Bigl (zI -L^{({\rm oo})}\Bigr )
\det \Bigl (zI -L^{({\rm ee})}-L^{({\rm eo})}
(zI-L^{({\rm oo})})^{-1}L^{({\rm oe})}\Bigr ).
$$
It is easy to see that the right 
hand side is finite as $\varepsilon \to 0$. In order to find the limit
as $\varepsilon \to 0$ we can put $L_{ij}^{({\rm oo})}=0$ and forget about
the next-to-leading powers of $\varepsilon$ in other blocks. 
In this way we find:
$$
\lim_{\varepsilon \to 0}\Bigl (L^{({\rm eo})}
(zI-L^{({\rm oo})})^{-1}L^{({\rm oe})}\Bigr )_{ij}=
\sigma (2\eta )z^{-1}U_i^- \phi (X_{ij}-2\eta , \lambda ).
$$
Therefore, the generating function of integrals of motion is
\beq\label{g9}
R(z, \lambda )=\det_{1\leq i,j\leq N}
\Bigl (z\delta_{ij}-\dot X_i \phi(X_{ij}\! -\! \eta , \lambda )-
\sigma (2\eta )z^{-1} U_i^- \phi(X_{ij}\! -\! 2\eta , \lambda )\Bigr ).
\eeq

\begin{proposition}
The generating function $R(z, \lambda )$ is given by
\beq\label{g10}
\begin{array}{l}
\displaystyle{
R(z, \lambda )=z^N +z^{-N}
\frac{\sigma (\lambda -2N\eta )}{\sigma (\lambda )}}
\\ \\
\displaystyle{\phantom{aaaaaaaaaaaaaaa}+\, 
\sum_{k=1}^N z^{N-k} \frac{\sigma (\lambda -k\eta )}{\sigma (\lambda )
\sigma (k\eta )}\, J_k+
\sum_{k=1}^{N-1} z^{k-N} 
\frac{\sigma (\lambda -2N\eta +k\eta )}{\sigma (\lambda )
\sigma (k\eta )}\, J_{-k}},
\end{array}
\eeq
where the integrals of motion are
$$
J_{\pm k}=\sum_{m=0}^{[k/2]}J_{\pm k, m}
$$
and $J_{\pm k, m}$ are given in (\ref{im7}).
\end{proposition}

\noindent
{\it Sketch of proof.} The proof is a lengthy but straightforward calculation
which uses the formula for determinant of sum of two matrices and
the formula for determinant of the elliptic Cauchy matrix (\ref{g4}).
Here are some details. First of all, the determinant $\det (I+M)$ is equal
to the sum of all diagonal minors of the matrix $M$ of all sizes,
including the ``empty minor'' which is put equal to $1$. After that
we encounter the determinants of the form
$\det (A_{\cal J} +B_{{\cal J}})$, where $A_{\cal J}$, $B_{{\cal J}}$
are diagonal minors of the 
matrices $\dot X_i \phi(X_{ij}\! -\! \eta , \lambda )$,
$\sigma (2\eta )z^{-1} U_i^- \phi(X_{ij}\! -\! 2\eta , \lambda )$
of size $n\leq N$ with rows and columns indexed by indexes from 
a set ${\cal J}=\{j_1, \ldots , j_n\}\subseteq \{1, \ldots , N\}$
($j_1< j_2 <\ldots <j_n \leq N$). The formula for 
determinant of sum of two matrices states that
$$
\det (A_{\cal J} +B_{{\cal J}})=\sum_{{\cal I}\subseteq {\cal J}}
\det A^{(B)}_{{\cal J}\setminus {\cal I}},
$$
where summation is carried out over all subsets ${\cal I}$ of the set
${\cal J}$ and $A^{(B)}_{{\cal J}\setminus {\cal I}}$ is the matrix 
$A_{\cal J}$
in which rows numbered by indexes from the set ${\cal I}$ are
substituted by the corresponding rows of the matrix $B_{{\cal J}}$.
Each $A^{(B)}_{{\cal J}\setminus {\cal I}}$ is an elliptic Cauchy matrix
(multiplied by a diagonal matrix), so the determinant of it is known.
To see this, we choose in (\ref{g4}) $x_j=X_j$ and
$$
\begin{array}{l}
y_j =X_j -\eta \;\;\; \mbox{if $j\in {\cal J}\setminus {\cal I}$},
\\ \\
y_j =X_j -2\eta \;\;\; \mbox{if $j\in {\cal I}$}.
\end{array}
$$
The determinant in (\ref{g9}) is represented as a Laurent
polynomial in $z$ with coefficients which are sums over sets
${\cal I}, \, {\cal I}'\subseteq \{1, \ldots , N\}$ 
such that ${\cal I}\cap {\cal I}'=\emptyset$ as in (\ref{im7}).
\square

The characteristic equation
\beq\label{g11}
R(z, \lambda )=0
\eeq
defines a Riemann surface $\tilde \Gamma$ which is a $2N$-sheet covering
of the $\lambda$-plane. This Riemann surface is an integral of motion. 
Any point of it is
$P=(z, \lambda )$, where $z, \lambda$ are connected 
by equation (\ref{g11}). There are $2N$
points above each point $\lambda$. 
It is easy to see from the right hand side of
(\ref{g10})
that the Riemann surface $\tilde \Gamma$ 
is invariant under the simultaneous transformations
\beq\label{trans}
\lambda \mapsto \lambda +2\omega, \quad z\mapsto 
e^{-2\zeta (\omega )\eta}z \quad \mbox{and} \quad
\lambda \mapsto \lambda +2\omega ', \quad z\mapsto 
e^{-2\zeta (\omega ')\eta}z.
\eeq
The factor of $\tilde \Gamma$ over the transformations (\ref{trans}) is an
algebraic curve $\Gamma$ which covers the elliptic curve with 
periods $2\omega , 2\omega '$.
It is the spectral curve of the deformed RS model.

\begin{proposition}
The spectral curve $\Gamma$ admits a holomorphic involution 
$\iota$ with two fixed points.
\end{proposition}

\noindent
{\it Proof.} In the previous section it was proved that $J_{-k}=J_k$.
Therefore, the equation $R(z, \lambda )=0$ is invariant under the
involution
\beq\label{g12}
\iota : (z, \lambda )\mapsto (z^{-1}, 2N\eta -\lambda ),
\eeq
as is easily seen from (\ref{g10}). The fixed points lie 
above the points $\lambda_{*}$ such that
$\lambda_{*} =2N\eta -\lambda_{*}$ modulo the lattice with 
periods $2\omega , 2\omega '$, i.e.
$\lambda_{*}=N\eta -\omega _{\alpha}$, where 
$\omega_{\alpha}$ is either $0$ or one of the three
half-periods $\omega_1=\omega$, $\omega_2 =\omega '$, 
$\omega_3=\omega +\omega '$. 
Substituting this into the equation of the spectral curve
and taking into account that $J_{-k}=J_k$, we conclude that
the fixed points are $(\pm 1, N\eta )$ and there are no fixed
points above $\lambda_{*}=N\eta -\omega _{\alpha}$ 
with $\omega_{\alpha}\neq 0$.
\square

\section{Conclusion and open problems}

In this paper we have found the complete set of integrals of motion
for the deformed RS system with equations of motion (\ref{int4}).
This provides enough evidence for integrability of the system. Our method
was based on the fact that the deformed RS system is equivalent to the
dynamical system for pairs of particles in the standard RS model (with
even number of particles) moving as whole things so that the distance
between particles in each pair is equal to $\eta$, the inverse ``velocity
of light'' in the RS ($=$ relativistic CM) model. Such pairs are
preserved by only a ``half'' of the higher Hamiltonian flows, so we 
consider only ${\sf H}_k^-$-flows and put the time variables 
associated with the ${\sf H}_k^+$-flows to zero. The configurations
in the full phase space ${\cal F}$ when particles stick together
in pairs form a half-dimensional subspace ${\cal P}\subset {\cal F}$
and we have proved that this subspace is Lagrangian and 
invariant under all
${\sf H}_k^-$-flows. Then integrals of motion for the deformed RS
system can be obtained by restricting the known RS integrals of motion
to the subspace ${\cal P}$. This job has been done in the present paper.

In the $\eta \to 0$ limit (in which the RS system reduces to the CM
system) the particles in each pair turn out to merge in one and the
same point. This singular limiting case was discussed in \cite{Z21}.

It is an interesting question whether any clusters of RS particles
other than pairs are possible in this sense. For example, one may
consider ``strings'' of $M$ particles such that the coordinates 
of the particles in the $i$th string are 
$x_{Mi+1-\alpha}=X_i+(M-\alpha )\eta$, $\alpha =1, \ldots , M$,
with $X_i$ being the coordinate of the string moving as a whole thing. 
It is natural to ask whether some Hamiltonian flows of the RS model
preserve such string structure.

We should stress that the connection between the standard RS system
and the deformed RS system is not trivial and has different aspects.
On one hand, the latter is an extension of the former and includes it
as a particular case because equations of motion (\ref{int4}) differ
from equations of motion (\ref{int2}) of the RS system by presence 
of some additional terms. However, on the other hand, the deformed
RS system is contained in the RS system since it can be regarded as
its reduction in the sense that the equations of motion (\ref{int4})
are obtained by restriction of the RS dynamics to the subspace ${\cal P}$
of pairs.

Finally, let us list some open problems which arise in connection with 
the deformed RS system. First, it is important to answer the question
whether the deformed RS system is Hamiltonian or not. A related problem
is quantization of the deformed RS system. Second, it would be highly
desirable to find a commutation representation for equations of motion
(\ref{int4}) such as Lax representation or Manakov's triple representation
\cite{Manakov}. It is the latter that is known to exist for equations
(\ref{int4a}) which can be obtained from (\ref{int4}) in the 
$\eta \to 0$ limit. That is why it is natural to conjecture that 
Manakov's triple representation exists for equations (\ref{int4})
for all $\eta \neq 0$.
Third, it would be interesting to find
B\"acklund transformations of the deformed RS system which are closely
connected with the so-called self-dual form of equations of 
motion and integrable time discretization of them. All this is known to
exist for the CM and RS systems (see \cite{W82}-\cite{Z19}).
We hope to discuss these problems elsewhere.

\section*{Appendix A: The Weierstrass functions}
\addcontentsline{toc}{section}{Appendix A: The Weierstrass functions}
\def\theequation{A\arabic{equation}}
\def\theHequation{\theequation}
\setcounter{equation}{0}

In this appendix we present the definition and main properties of the 
Weierstrass functions: 
the $\sigma$-function, the $\zeta$-function and the $\wp$-function
which are widely used in the main text.

Let $\omega$, $\omega '$ be complex numbers such that 
${\rm Im} (\omega '/ \omega )>0$.
The Weierstrass $\sigma$-function 
with quasi-periods $2\omega$, $2\omega '$ 
is defined by the following infinite product over the lattice
$2\omega m+2\omega ' m'$, $m,m'\in \ZZ$:
\beq\label{A1}
\sigma (x)=\sigma (x |\, \omega , \omega ')=
x\prod_{s\neq 0}\Bigl (1-\frac{x}{s}\Bigr )\, 
e^{\frac{x}{s}+\frac{x^2}{2s^2}},
\quad s=2\omega m+2\omega ' m' \quad m, m'\in \ZZ .
\eeq 
It is an odd entire quasiperiodic function in the complex plane. 
As $x\to 0$,
\beq\label{A1a}
\sigma (x)=x+O(x^5), \quad x\to 0.
\eeq
The monodromy properties of the $\sigma$-function 
under shifts by the quasi-periods
are as follows:
\beq\label{A4}
\begin{array}{l}
\sigma (x+2\omega )=-e^{2\zeta (\omega )(x+\omega )}\sigma (x),
\\ \\
\sigma (x+2\omega ' )=-e^{2\zeta (\omega ')(x+\omega ' )}\sigma (x).
\end{array}
\eeq
Here $\zeta (x)$ is the
Weierstrass $\zeta$-function defined as
\beq\label{A4a}
\zeta (x)=\frac{\sigma '(x)}{\sigma (x)}.
\eeq
The monodromy properties imply that the function
$$
f(x)=\prod_{\alpha =1}^M \frac{\sigma 
(x-a_{\alpha})}{\sigma (x-b_{\alpha})}, \qquad \sum_{\alpha =1}^M
(a_{\alpha}-b_{\alpha})=0
$$
is a double-periodic function with periods $2\omega$, $2\omega '$
(an elliptic function).

The Weierstrass $\zeta$-function
can be represented as a sum over the lattice as follows:
\beq\label{A2}
\zeta (x)=
\frac{1}{x}+\sum_{s\neq 0} \Bigl ( \frac{1}{x-s}+\frac{1}{s}+
\frac{x}{s^2}\Bigr ),
\quad s=2\omega m+2\omega ' m' \quad m, m'\in \ZZ .
\eeq
It is an odd function with first order poles at the points 
of the lattice. As $x\to 0$,
\beq\label{A2a}
\zeta (x)=\frac{1}{x} +O(x^3), \quad x\to 0.
\eeq
If the argument
is shifted by any quasi-period, the $\zeta$-function is transformed as
follows:
\beq\label{A5}
\begin{array}{l}
\zeta (x+2\omega )=\zeta (x)+\zeta (\omega ),
\\ \\
\zeta (x+2\omega ' )=\zeta (x)+\zeta (\omega ').
\end{array}
\eeq
These values $\zeta (\omega )$, $\zeta (\omega ')$
are related by the identity
$2\omega ' \zeta (\omega )-2\omega \zeta (\omega ')=\pi i$.
The transformation properties (\ref{A5}) imply that the function
$$
g(x)=\sum_{\alpha =1}^M A_{\alpha} \zeta (x-a_{\alpha}), \qquad
\sum_{\alpha =1}^M A_{\alpha}=0
$$
is an elliptic function. 

The Weierstrass $\wp$-function is defined as
$\wp (x)=-\zeta '(x)$. It can be represented as a sum over the 
lattice as follows:
\beq\label{A3}
\wp (x)= \frac{1}{x^2}+\sum_{s\neq 0} \Bigl ( 
\frac{1}{(x-s)^2}-\frac{1}{s^2}\Bigr ),
\quad s=2\omega m+2\omega ' m' \quad m, m'\in \ZZ .
\eeq
It is an even double-periodic function with periods $2\omega , 2\omega '$
and with second order poles at the points 
of the lattice $s=2\omega m+2\omega ' m'$ with integer $m, m'$.
As $x\to 0$, $\wp (x)=x^{-2}+O(x^2)$.

\section*{Appendix B: Proof of Lemma \ref{lemma:identity}}
\addcontentsline{toc}{section}{Appendix B: 
Proof of Lemma \ref{lemma:identity}}
\def\theequation{B\arabic{equation}}
\def\theHequation{\theequation}
\setcounter{equation}{0}

We set
\beq\label{B1}
F_m^{\pm}=
\sum_{{\cal I}\subset {\cal N}'\atop |{\cal I}|=m}
\prod_{i\in {\cal I}}\prod_{\ell 
\in {\cal N}'\setminus {\cal I}}U^{\pm}(X_{i\ell})=
\sum_{{\cal I}\subset {\cal N}'\atop |{\cal I}|=m}
\prod_{i\in {\cal I}}\prod_{\ell 
\in {\cal N}'\setminus {\cal I}}\frac{\sigma (X_{i\ell}\pm 2\eta )\,
\sigma (X_{i\ell}\mp \eta )}{\sigma (X_{i\ell}\pm \eta )\,
\sigma (X_{i\ell})}
\eeq
and consider the function $f_m=F_m^{+}-F_m^{-}$. It is a symmetric 
function of the variables $X_j$, $j\in {\cal N}'$. It is easy to see that
it is an elliptic function of each $X_j$. 
The statement of the lemma is that $f_m=0$ for all $m$.
At $m=1$ we have:
\beq\label{B2}
f_1=\sum_{i\in {\cal N}'}\prod_{\ell 
\in {\cal N}'\atop \ell \neq i}
\frac{\sigma (X_{i\ell}+ 2\eta )\,
\sigma (X_{i\ell}- \eta )}{\sigma (X_{i\ell}+ \eta )\,
\sigma (X_{i\ell})}-
\sum_{i\in {\cal N}'}\prod_{\ell 
\in {\cal N}'\atop \ell \neq i}
\frac{\sigma (X_{i\ell}- 2\eta )\,
\sigma (X_{i\ell}+ \eta )}{\sigma (X_{i\ell}- \eta )\,
\sigma (X_{i\ell})}=0
\eeq
since it is proportional to 
the sum of residues of the elliptic function
$$
f(X)=\prod_{\ell \in {\cal N}'}\frac{\sigma (X-X_{\ell}+2\eta )\, 
\sigma (X-X_{\ell}-\eta )}{\sigma (X-X_{\ell}+\eta )\,
\sigma (X-X_{\ell})}.
$$

We are going to prove that $f_m=0$ for all $m$ by induction. 
Suppose that $f_m=0$ for some $m$; we will show that this is also
true for $m\to m+1$. Due to the symmetry, it is 
enough to consider $f_m$ as a function of $X_1$ (without loss of generality
we assume that ${\cal N}'\ni 1$). Possible poles of this function 
are first order poles at $X_1=X_j$ and $X_1=X_j\pm \eta$. Let us prove
that residues at these poles actually vanish. For the poles at
$X_1=X_j$ this is especially simple  because it is not difficult to
see that $\displaystyle{\res_{X_1=X_j}F_m^{\pm}=0}$ even without 
the inductive assumption. Consider the pole at $X_1=X_2+\eta$
(again, without loss of generality
we can assume that ${\cal N}'\ni 2$). Let us introduce the short-hand
notation ${\cal N}'_{1}={\cal N}'\setminus \{1\}$, 
${\cal N}'_{2}={\cal N}'\setminus \{2\}$, 
${\cal N}'_{12}={\cal N}'\setminus \{1,2\}$. Then we have:
\beq\label{B3}
\begin{array}{lll}
\res_{X_1=X_2+\eta} f_m &=&\displaystyle{\sigma (2\eta )
\sum_{{\cal I}\subseteq {\cal N}'_{12}}
\prod_{\ell \in {\cal N}'_{12}\setminus {\cal I}}
U^-(X_{1\ell}) \prod_{i\in {\cal I}}\prod_{\ell \in {\cal N}'_{1}
\setminus {\cal I}}
U^-(X_{i\ell})}
\\ && \\
&&\phantom{aaaaaa}- \, \displaystyle{\sigma (2\eta )
\sum_{{\cal I}\subseteq {\cal N}'_{12}}
\prod_{\ell \in {\cal N}'_{12}\setminus {\cal I}}
U^+(X_{2\ell}) \prod_{i\in {\cal I}}\prod_{\ell \in {\cal N}'_{2}
\setminus {\cal I}}
U^+(X_{i\ell})},
\end{array}
\eeq
where $|{\cal I}|=m-1$. Since 
$X_1=X_2+\eta$, we have $U^+(X_{2\ell})=U^-(X_{1\ell})$. After simple
transformations of the products, we can represent (\ref{B3}) in the form
\beq\label{B4}
\begin{array}{l}
\displaystyle{
\res_{X_1=X_2+\eta} f_m=\sigma (2\eta )\prod_{\ell \in {\cal N}'_{12}}
\! U^-(X_{1\ell})\left [
\sum_{{\cal I}\subseteq {\cal N}'_{12}}
\prod_{i\in {\cal I}}\prod_{\ell \in {\cal N}'_{12}
\setminus {\cal I}}\! U^-(X_{i\ell})
-\! \sum_{{\cal I}\subseteq {\cal N}'_{12}}
\prod_{i\in {\cal I}}\prod_{\ell \in {\cal N}'_{12}
\setminus {\cal I}}\! U^+(X_{i\ell})\right ]}
\end{array}
\eeq
The expression in the square brackets is nothing else than 
$f_{m-1}$ which is zero by the induction assumption. Therefore,
$\displaystyle{\res_{X_1=X_2+\eta} f_m=0}$ for all $m$. The pole 
at $X_1=X_2-\eta$ and the poles at $X_1=X_j\pm \eta$
are considered in the similar way. We have shown
that the elliptic function $f_m$ as a function of $X_1$ is regular.
Therefore, it does not depend on $X_1$. 
By virtue of the symmetry, this function is a constant which
does not depend on all $X_j$'s. To find the constant, one may put
$X_{j}=\varepsilon j$ and tend $\varepsilon \to 0$. It is easy to
see that $f_m$ after this substitution 
is an odd function of $\varepsilon$, so the constant term $\propto
\varepsilon^0$ in the expansion as $\varepsilon \to 0$ vanishes.
This means that the constant is equal to zero. 

\section*{Acknowledgments}

\addcontentsline{toc}{section}{Acknowledgments}

I thank A.Marshakov and V.Prokofev for discussions. 
This work has been supported in part within the framework of the
HSE University Basic Research Program.

\end{document}